\documentclass[a4paper,11pt]{article}
\usepackage{jcappub} 
\usepackage{lineno}
\usepackage[nameinlink,noabbrev]{cleveref}
\usepackage{xspace}
\usepackage{xcolor}
\usepackage{url}
\usepackage{aas_macros}
\graphicspath{{fig/}}
\usepackage{scalerel}
\usepackage{hanging}

\usepackage{orcidlink}

\title{\boldmath DESI 2024: Reconstructing Dark Energy using Crossing Statistics with DESI DR1 BAO data}

\newcommand{\fde}{\ensuremath{f_{\rm DE}}}

\newcommand{\chisq}{\ensuremath{\chi^2}}
\newcommand{\dchisq}{\ensuremath{\Delta\chi^2}}

\newcommand{\lcdm}{$\Lambda$CDM}

\newcommand{\lya}{Ly$\alpha$\xspace}

\newcommand{\Omo}{\ensuremath{\Omega_{\text{m},0}}}

\newcommand{\ocdm}{\ensuremath{\omega_\text{cdm}}}
\newcommand{\ob}{\ensuremath{\omega_\text{b}}}

\newcommand{\zrec}{\ensuremath{z_\text{rec}}}
\newcommand{\DVrd}{D_\mathrm{V}/r_\mathrm{d}}
\newcommand{\DMrd}{D_\mathrm{M}/r_\mathrm{d}}
\newcommand{\DHrd}{D_\mathrm{H}/r_\mathrm{d}}

\newcommand{\Hord}{\ensuremath{H_0\rd}}

\newcommand{\rd}{\ensuremath{r_\text{d}}} 
\newcommand{\diff} {\ensuremath{\mathrm{d}}}

\newcommand{\classy}{\texttt{class}}
\newcommand{\Cobaya}{\texttt{cobaya}}
\newcommand{\gd}{\texttt{GetDist}}
\newcommand{\kmsMpc}{\,{\rm km\,s^{-1}\,Mpc^{-1}}}
\newcommand{\kms}{\,{\rm km\,s^{-1}}}

\crefname{equation}{Eq.}{Eqs.}
\crefname{section}{Section}{Sections}
\crefname{figure}{Fig.}{Figs.}
\crefname{table}{Table}{Tables}
\crefname{appendix}{Appendix}{Appendices}
\Crefname{figure}{Figure}{Figures}
\Crefname{equation}{Equation}{Equations}
\Crefname{section}{Section}{Sections}
\Crefname{table}{Table}{Tables}

\def\tpdf#1{\texorpdfstring{#1}{Lg}}

\definecolor{llgray}{gray}{0.93}
\definecolor{lgray}{gray}{0.83}
\definecolor{deepmagenta}{rgb}{0.8, 0.0, 0.8}
\definecolor{ballblue}{rgb}{0.13, 0.67, 0.8}
\definecolor{celestialblue}{rgb}{0.29, 0.59, 0.82}
\definecolor{RedWine}{rgb}{0.743,0,0}
\definecolor{DarkGreen}{rgb}{0,0.6,0}

\emailAdd{calderon@kasi.re.kr, shafieloo@kasi.re.kr}
\affiliation{Affiliations are in Appendix \ref{sec:affiliations}}

\author[1]{{R.~Calderon}\orcidlink{0000-0002-8215-7292},}
\author[1,2]{{K.~Lodha}\orcidlink{0009-0004-2558-5655},}
\author[1,2]{{A.~Shafieloo}\orcidlink{0000-0001-6815-0337},}
\author[3,4,5]{{E.~Linder}\orcidlink{0000-0001-5536-9241},}
\author[1]{{W.~Sohn},}
\author[6]{{A.~de~Mattia},}
\author[7]{{J.~L.~Cervantes-Cota}\orcidlink{0000-0002-3057-6786},}
\author[8]{{R.~Crittenden},}
\author[9]{{T.~M.~Davis}\orcidlink{0000-0002-4213-8783},}
\author[10]{{M.~Ishak}\orcidlink{0000-0002-6024-466X},}
\author[3]{{A.~G.~Kim}\orcidlink{0000-0001-6315-8743},}
\author[1]{{W.~Matthewson},}
\author[11,12]{{G.~Niz}\orcidlink{0000-0002-1544-8946},}
\author[1,2]{{S.~Park}\orcidlink{0009-0000-2266-9985},}
\author[3]{{J.~Aguilar},}
\author[13]{{S.~Ahlen}\orcidlink{0000-0001-6098-7247},}
\author[14,15]{{S.~Allen},}
\author[16]{{D.~Brooks},}
\author[3]{{T.~Claybaugh},}
\author[17]{{A.~de la Macorra}\orcidlink{0000-0002-1769-1640},}
\author[18]{{A.~Dey}\orcidlink{0000-0002-4928-4003},}
\author[19]{{B.~Dey}\orcidlink{0000-0002-5665-7912},}
\author[16]{{P.~Doel},}
\author[20,21]{{J.~E.~Forero-Romero}\orcidlink{0000-0002-2890-3725},}
\author[22,8,23]{{E.~Gaztañaga},}
\author[3]{{S.~Gontcho A Gontcho}\orcidlink{0000-0003-3142-233X},}
\author[24,25,26]{{K.~Honscheid},}
\author[9]{{C.~Howlett}\orcidlink{0000-0002-1081-9410},}
\author[18]{{S.~Juneau},}
\author[3]{{A.~Kremin}\orcidlink{0000-0001-6356-7424},}
\author[3]{{M.~Landriau}\orcidlink{0000-0003-1838-8528},}
\author[27]{{L.~Le~Guillou}\orcidlink{0000-0001-7178-8868},}
\author[3]{{M.~E.~Levi}\orcidlink{0000-0003-1887-1018},}
\author[28,29]{{M.~Manera}\orcidlink{0000-0003-4962-8934},}
\author[30,29]{{R.~Miquel},}
\author[31]{{J.~Moustakas}\orcidlink{0000-0002-2733-4559},}
\author[19]{{J.~ A.~Newman}\orcidlink{0000-0001-8684-2222},}
\author[6,3]{{N.~Palanque-Delabrouille}\orcidlink{0000-0003-3188-784X},}
\author[32,33,34]{{W.~J.~Percival}\orcidlink{0000-0002-0644-5727},}
\author[3,4,5]{{C.~Poppett},}
\author[35]{{F.~Prada}\orcidlink{0000-0001-7145-8674},}
\author[36]{{M.~Rezaie}\orcidlink{0000-0001-5589-7116},}
\author[37]{{G.~Rossi},}
\author[6]{{V.~Ruhlmann-Kleider}\orcidlink{0009-0000-6063-6121},}
\author[38]{{E.~Sanchez}\orcidlink{0000-0002-9646-8198},}
\author[3]{{D.~Schlegel},}
\author[39,40]{{M.~Schubnell},}
\author[41]{{H.~Seo}\orcidlink{0000-0002-6588-3508},}
\author[18]{{D.~Sprayberry},}
\author[40]{{G.~Tarl\'{e}}\orcidlink{0000-0003-1704-0781},}
\author[26]{{P.~Taylor},}
\author[17]{{M.~Vargas-Maga\~na}\orcidlink{0000-0003-3841-1836},}
\author[18]{{B.~A.~Weaver},}
\author[27]{{P.~Zarrouk}\orcidlink{0000-0002-7305-9578},}
\author[42]{{H.~Zou}\orcidlink{0000-0002-6684-3997},}

\abstract{We implement Crossing Statistics to reconstruct in a model-agnostic manner the expansion history of the universe and properties of dark energy, using DESI Data Release 1 (DR1) BAO data in combination with one of three different supernova compilations (PantheonPlus, Union3, and DES-SN5YR) and Planck CMB observations. Our results hint towards an evolving and emergent dark energy behaviour, with negligible presence of dark energy at $z\gtrsim 1$, at varying significance depending on data sets combined. In all these reconstructions, the cosmological constant lies outside the 95\% confidence intervals for some redshift ranges. This dark energy behaviour, reconstructed using Crossing Statistics, is in agreement with results from the conventional $w_0$--$w_a$ dark energy equation of state parametrization reported in the DESI Key cosmology paper. Our results add an extensive class of model-agnostic reconstructions with acceptable fits to the data, including models where cosmic acceleration slows down at low redshifts. We also report constraints on \Hord\ from our model-agnostic analysis,  
independent of the pre-recombination physics. 
}

\begin{document}
\maketitle
\flushbottom

\section{Introduction}
The concordance model of cosmology, rooted in Einstein's general theory of relativity (GR), provides a testable and falsifiable theory of the Universe. As an isotropic and homogeneous universe, with standard model radiation and matter, plus cold dark matter and a cosmological constant, the concordance \lcdm\ model has proved to be highly successful in explaining the wide array of cosmological observations~\cite{SupernovaSearchTeam:1998fmf,SupernovaCosmologyProject:1998vns,Planck1:2020,Alam_2021,Zhao_2022}. 
This predictive model effectively describes the Universe's dynamics with only 6 free parameters. However, cosmic acceleration has such fundamental implications, that a key aim in physical cosmology has been to challenge the assumption that dark energy is synonymous with the cosmological constant \cite{Ostriker:1995su, RevModPhys.61.1,Sahni:1999gb,Peebles:2002gy,Copeland:2006wr,Bull:2015stt,Perivolaropoulos:2021jda}, a query central to many cosmological surveys and instruments \cite{DES:2017myr,DES:2017qwj,eBOSS:2020yzd,Heymans:2020gsg,DES:2021wwk,DESI2024.I.DR1}. Many cosmologists have aimed to reconstruct the expansion history of the universe and understand the properties of dark energy, leading to the introduction of numerous parametric and nonparametric approaches~\citep{Huterer:2002hy,Shafieloo:2005nd,2007MNRAS.380.1573S,2009JCAP...12..025C,Bogdanos:2009ib,2010PhRvD..82j3502H,PhysRevLett.105.241302,Nesseris:2010ep,Shafieloo:2012ht,2012PhRvL.109q1301Z,Zhao:2012aw,Nesseris_2012,2013PhRvD..87b3520S,LHuillier:2016mtc,Koo:2020wro,Calderon:2022cfj,Calderon2023}. 

In this study, we employ the technique of Crossing Statistics \cite{Shafieloo:2012jb,2012JCAP...08..002S} on DESI BAO data, combined with supernovae and CMB data, to reconstruct the expansion history of the universe and properties of dark energy. 
Crossing Statistics serve to assess deviations from an assumed background model (in this case, the cosmological constant as dark energy) and offer a wide range of viable reconstructions tailored to different data combinations. Our findings suggest that the data combination is consistent with, and to some extent in favor of, an evolving emergent behaviour of dark energy with a slowing down of the acceleration at low redshifts.

\section{Methodology and Data}

In this paper, we use the most recent Baryon Acoustic Oscillations (BAO) observations from the Dark Energy Spectroscopic Instrument (DESI) along with different supernova compilations and cosmic microwave background data. Here we briefly explain the methods and data we use in this analysis.

\subsection{Crossing Statistics}

Crossing Statistics were originally introduced to extract information from data beyond the conventional $\chi^2$ statistics and its relation to likelihoods. With the median statistic as its first mode and the $\chi^2$ statistic as its last mode, Crossing Statistics can be considered as a generalization of $\chi^2$ statistics~\citep{Shafieloo:2010xm}. While the original Crossing Statistics are hard to implement for the purpose of model validation, model selection, and parameter estimation, their Bayesian interpretation can be trivially used for such purposes~\citep{Shafieloo:2012jb,2012JCAP...08..002S}. Crossing Statistics can also be used in some other contexts, including reconstruction and consistency tests~\citep[see also \cite{Haude:2019qms} for the use of Chebyshev polynomials for reconstruction.]{Hazra:2013oqa,Hazra:2014hma,Shafieloo:2016zga}.
\\

In this work, we use the Crossing Statistic formalism for the reconstruction of the dark energy properties \cite{Shafieloo:2012jb,2012JCAP...08..002S}. In particular, we use the fact that any given function $y(z)$ can be expanded as a Chebyshev series
\begin{equation}\label{eq:Chebyshev}
     y(z)=\sum_{i=0}^{N} C_i ~T_i(x), \qquad
    x\equiv1 -2\,\frac{z_\mathrm{max}-z} {z_\mathrm{max}-z_\mathrm{min}} \in [-1,1]~,
\end{equation}
where $T_i(x)$ are the Chebyshev polynomials of the first kind, defined in the interval $x\in[-1,1]$, and $C_i$ are some coefficients multiplying these polynomials. The Chebyshev polynomials, in the limit $N\rightarrow\infty$, form a basis that spans the space of all continuous functions.
For practical purposes, we limit the analysis to four terms in the expansion; including higher orders leads to weaker constraints, due to the increased number of degrees of freedom\footnote{In practice, the maximum order of polynomials should be set as the smallest integer compatible with the data. Previous works have shown that using up to four terms of Crossing Statistics is generally sufficient to capture the time evolution of smooth functions in a limited range \cite{Shafieloo:2012jb,Hazra:2013oqa}.}. In \cref{exp-order} we provide further motivations for this choice.
The first four Chebyshev polynomials are given by
\begin{equation}
\label{eq:polynomials}
\begin{aligned}
T_0(x) & = 1,
\qquad
T_1(x)  = x \,,\\
T_2(x) & = 2x^2 - 1\,,
\qquad
T_3(x) =  4x^3-3x \,.
\end{aligned}
\end{equation}
We use these polynomials to model the energy density and equation of state of dark energy and to reconstruct its possible time-dependence from the data, as described in \cref{sec:DE} below. We choose $z_{\rm min}=0$ and $z_{\rm max}=3.5$, 
beyond which redshift there are no more data to constrain its evolution. See the individual subsections below regarding $z>z_{\rm max}$. 
We have verified that the results do not differ significantly if instead we choose $z_{\rm max}=2.35$ (see \Cref{sec:zmax}). 
Note that many stage 4 surveys will probe physics at $z\lesssim 3.5$ \citep{LSST,Euclid:2021icp,2009pra..confE..58L,Wang:2021oec}. Also note that unlike previous analyses using crossing statistics \citep{Hazra:2013oqa,Hazra:2014hma,Shafieloo:2016zga}, where $x\in [0,1]$ was used, we use the full range $x\in[-1,1]$ spanned by the polynomials 
(cf.\ \cref{eq:Chebyshev}).\\

In the Bayesian interpretation of the crossing statistic \cite{2012JCAP...08..002S}, the posterior distribution of the Chebyshev coefficients $C_i$ carries valuable information on the model (mean function) at hand. Any significant deviation from the mean function $C_i$'s would indicate that the data favors deformations away from the considered model (see \cref{sec:hyperapx}).

\subsection{Dark energy modeling}\label{sec:DE}
Assuming a flat universe, the Friedmann equation reads
\begin{equation}\label{eq:h2}
    \frac{H^2(z)}{H_0^2}=\Omega_{\rm m,0}(1+z)^3+\Omega_{\rm r,0}(1+z)^4+(1-\Omega_{\rm m,0}-\Omega_{\rm r,0})\;f_{\rm DE}(z)
\end{equation}
where $\Omega_{i,0}\equiv 8\pi G\rho_i/3H_0^2$ are the fractional energy density parameters of the respective components at the present time and $f_{\rm DE}(z)\equiv\rho_{\rm DE}(z)/\rho_{\rm DE,0}$ is the effective (normalized) dark energy density, which accounts for any additional contribution to the expansion rate coming from e.g.\ unknown energy density or modified gravity. 
Using the conservation of energy, the energy density can be related to the pressure to energy density ratio of dark energy through its equation of state parameter $w(z)$, 
\begin{equation}
     \fde(z)\equiv\frac{\rho_{\rm DE}(z)}{\rho_{\rm DE,0}}  =\exp{\left(3\int_0^z\left[1+w(\tilde{z})\right]~\diff\ln{(1+\tilde{z})}\right)}\ .
     \label{eqn_fde}
\end{equation}

A common approach used to explore alternative dark energy models is to parameterize the time evolution of its equation of state. The standard parametrization, $w(a) =w_0+w_a\,(1 - a)$ \cite{2001IJMPD..10..213C,Linder:2002et}, has been shown to be highly accurate for a wide variety of models \citep{Linder:2002et,dePutter:2008wt}.\\

Alternatively, a bottom-up approach involves reconstructing its redshift evolution directly from the data using non-parametric methods (see for example \cite{Huterer:2002hy,Shafieloo:2005nd,PhysRevLett.105.241302,Nesseris_2012,Shafieloo:2007cs,Shafieloo:2012ht,2012PhRvL.109q1301Z,2013PhRvD..87b3520S,LHuillier:2016mtc, Zhao:2017cud, Poulin:2018zxs, Wang:2018fng, Keeley:2019esp,LHuillier:2019imn,Koo:2020wro, Calderon:2022cfj, Calderon2023, Raveri_2023}). In this work, we explore an extended reconstruction for dark energy, both in terms of its equation of state and its energy density to see if the data suggest more complicated dynamics.  

\subsubsection{Equation of state}\label{sec:EoS}

We start by exploring a flexible parameterization for the equation of state, $w(z)$. We use the Chebyshev expansion in \cref{eq:Chebyshev} with four terms, around $w=-1$, 
\begin{equation}
     w(z)  = -1 \times \sum_{i=0}^{N=3} C_i ~T_i(x)~.
\end{equation}
This parametrization introduces four additional degrees of freedom, captured by the four coefficients $C_i$ multiplying the polynomials given by \cref{eq:polynomials}. For moderate redshift (geometrical) probes, such as BAO and SN, we compute distances and constrain the energy content of our models, treating the degenerate combination $\Hord$ as a free parameter. We will refer to this as ``background-only''. When including the measurements from the cosmic microwave background, we implement this parametrization using our modified version of the Boltzmann solver \classy\ \cite{class_2011arXiv1104.2932L,class_2011JCAP...07..034B} that allows for an arbitrary redshift dependence of $w(z)$. 
In order that we give enough flexibility to the dark energy behaviour at low redshifts (where dark energy is most dominant) \textemdash and to be consistent with our ``background-only'' runs \textemdash beyond $z_{\rm max}=3.5$ we smoothly extrapolate the equation of state $w(z)$ to a cosmological constant $w=-1$  using a transition function $-1+(B_0 + B_1 \cdot u) e^{-u^2/\Delta^2}$ where $u = \log{\frac{1+z}{1+ z_{\rm max}}}$, and $B_0, B_1$ are chosen to ensure $w(z)$ remains smooth and differentiable at the transition. 
Unlike \cite{Keeley:2022ojz}, we give more freedom to the equation of state, focusing on the very low-$z$ regime and using the whole range spanned by the polynomials, whereas a very broad $x=\log(1+z)/\log(1+z_{\rm rec})$ was used in \cite{Keeley:2022ojz}. 
To allow for a phantom equation of state, $w(z)<-1$, we rely on the Parametrized-Post Friedmann (PPF) framework \cite{Hu:2007pj}, as implemented in \classy. Moreover, our main focus being the geometrical probes, all the runs in this analysis assume a sound speed $c_s^2=1$ for the dark energy fluid. Including dark energy perturbations is beyond the scope of this work. We present our findings, using both the ``background-only'' probes and those including the CMB, in \cref{sec:results_w}.

\subsubsection{Dark energy density}\label{sec:fde-model}
Alternatively to the equation of state, one can directly reconstruct the redshift dependence of the dark energy density.
The main advantage of working with the effective energy density, $\rho_{\rm DE}^{\rm eff}(z)$, is that it allows us to cover a larger class of models. Unlike the equation of state parameter $w(z)$, the direct reconstruction of $\fde(z)$ allows for the effective energy density to become negative, which can happen in modified gravity \citep{Sahni_2003,Zhou_2009,Bauer_2010,Matsumoto_2018,Tiwari_2024} and various other dark energy models \citep{Calderon:2020hoc,Vazquez:2012ag,Sahni:2014ooa,Sola:2015wwa,Lymperis_2022,Akarsu_2015,Ludwick_2017,Gomez-Valent:2024tdb}. The normalized energy density evolution $\fde$ is then written as
\begin{equation}\label{eq:fCi} 
    \fde(z) = C_0+\sum_{i=1}^{N=3} C_i ~T_i(x)~.
\end{equation} 
The function \fde, defined in \cref{eqn_fde}, satisfies by definition  $\fde(z=0)\equiv1$. Thus, $C_0$ is not a free-parameter and must be determined for each $C_{i>0}$ using the closure relation
\begin{equation}\label{eq:C0}
    C_0=1-\sum_{i=1}^{N=3} C_i ~T_i(x=-1)=1-\sum_{i=1}^{N=3}C_i\times (-1)^i~.
\end{equation}
Note that a cosmological constant ($\fde=1$) is recovered for $C_0=1,C_{i>0}=0$. Due to complications arising from the treatment of perturbations with negative energy densities, we restrict this part of the analysis to the ``background-only'' probes, and so always $z<z_{\rm max}$. 
The results when directly reconstructing $\fde(z)$ are presented in \cref{sec:negative_fde}.

\subsection{Data}\label{sec:data}

\begin{itemize}
    \item \textbf{Baryon Acoustic Oscillations (BAO):} 
    We use the compilation of compressed distance quantities $\DMrd$, $\DHrd$, and $\DVrd$ from the first year data release (DR1) of the Dark Energy Spectroscopic Instrument (DESI) \cite{DESI2016a.Science,DESI2022.KP1.Instr,DESI2023a.KP1.SV,DESI2023b.KP1.EDR,DESI2024.I.DR1}, as given in \cite{DESI2024.VI.KP7A}.  
    BAO measures effective distances relative to the drag-epoch sound horizon $r_d\equiv r_s(z_{\rm drag})$. Along the line of sight we measure 
    \begin{equation}
    \frac{D_H(z)}{r_d}\equiv\frac{c}{H(z)r_d}=\frac{c}{H_0r_d}\frac{1}{h(z)}\ ,
    \end{equation}
    where $h(z)=H(z)/H_0$, and transverse to the line of sight we measure   
    \begin{equation}
    \frac{D_M(z)}{r_d}\equiv\frac{c}{r_d}\int_0^z\frac{d\tilde{z}}{H(\tilde{z})}=\frac{c}{H_0r_d}\int_0^z\frac{d\tilde{z}}{h(\tilde{z})}\ ,
    \end{equation}
while $D_V\equiv [zD_M^2(z)D_H(z)]^{1/3}$ is an angle-averaged effective ``monopole'' distance. 
    This dataset, abbreviated as ``DESI BAO'', spans seven redshift bins from $z=0.3$ to $z=2.33$ (where the last bin extends out to $z\approx3.5$) \cite{DESI2024.III.KP4}. Additionally, in some cases (see \Cref{sec:SDSS}) we use the combination of DESI and SDSS data, referred to as ``(DESI+SDSS) BAO'', taking the three redshift bins at $z<0.6$ from SDSS with tighter constraints, while keeping DESI results for $z>0.6$. For \lya\ data, we employ the combined DESI+SDSS results provided by \cite{DESI2024.IV.KP6}. We direct the reader to \cite{DESI2024.I.DR1,DESI2024.II.KP3,DESI2024.III.KP4,DESI2024.V.KP5,DESI2024.IV.KP6,DESI2024.VI.KP7A,DESI2024.VII.KP7B} for further details on these data sets and BAO combination choices.
     
    \item \textbf{Supernovae Ia (SNe~Ia):} For some combined data sets we use supernova data from  
    three compilations, one at a time: ``PantheonPlus'', a compilation of 1550 supernovae spanning $0.01$ to $2.26$ \cite{Brout:2022vxf}; ``Union3'', containing 2087 SNe~Ia processed through the Unity 1.5 pipeline based on Bayesian Hierarchical Modelling \cite{Rubin:2023ovl}; and ``DES-SN5YR'', a compilation of 194 low-redshift SNe~Ia ($0.025<z<0.1$) and 1635 photometrically classified SNe~Ia covering the range $0.1<z<1.3$ \cite{DES:2024tys}. SNe~Ia give measures of luminosity distances $D_L(z)=(1+z)\,D_M(z)$. 
    
    \item \textbf{Cosmic Microwave Background (CMB):} We also include temperature and polarisation measurements of the CMB from the Planck satellite \cite{Planck:2018vyg}. In particular, we use the high-$\ell$ TTTEEE likelihood (\texttt{planck\_2018\_highl\_plik.TTTEEE}), together with low-$\ell$ TT (\texttt{planck\_2018\_lowl.TT}) and low-$\ell$ EE (\texttt{planck\_2018\_lowl.EE}) \cite{Aghanim:2019ame}, as implemented in \Cobaya~\cite{Torrado:2020dgo}. 
\end{itemize}

\subsection{Analysis}\label{sec:analysis}

We perform an MCMC sampling of the parameter space using the Metropolis-Hastings \cite{Lewis:2002ah,Lewis:2013hha} algorithm implemented in the publicly available sampler \Cobaya~\cite{Torrado:2020dgo}.  The priors used in the analysis are given in \cref{tab:priors}. When considering non-CMB observations (i.e.\ SNe and BAO), we use a custom theory code inheriting from \Cobaya's \texttt{Theory} base class to compute the observables (i.e.\ distances). This only requires the knowledge of the background expansion history (as implemented in \cref{eq:h2} above). For such cases, we sample $\vec{\theta}=\{
\Omo,C_0,C_1,C_2,C_3,\Hord\}$, where the $C_i$'s determine the dark energy behaviour and we treat the combination \Hord\ as a nuisance parameter. 
When including the CMB likelihood, we use our modified version of the Boltzmann solver \classy\ to implement a custom equation of state for dark energy. In that case, the parameter space associated with our model is $\vec{\theta}=\{
\ocdm,\ob,\ln{(10^{10} A_{s})},n_s,\tau,H_0,C_0,C_1,C_2,C_3\}$ where $r_d$ is no longer a free parameter, but its value is rather derived assuming standard pre-recombination physics. Throughout this work, we assume two massless and one massive neutrino, with $m_\nu=0.06 ~\rm eV$. We take advantage of the ``fast-dragging'' scheme \cite{Neal:2005} when sampling the CMB Planck likelihoods. For the PantheonPlus, DES-SN5YR and Union3 likelihoods, the marginalization over the absolute magnitude $M_B$ is done analytically.  
The details of the dark energy modelling and its numerical implementation are given in \cref{sec:DE}.

\begin{table}[t]
    \caption{
    Parameters and priors used in the analysis. All of the priors are uniform in the ranges specified below, except for the hyperparameters $C_i$ where we used a Gaussian prior centered around the mean $\mu^{\Lambda\rm CDM}$  ($C_0=1$, $C_{i>0}=0$).  
    We consider two sets of parameters, ``background-only'' when using BAO and SNe data only, and ``CMB'' where our modified version of the Boltzmann solver \classy\ has been used.}
    \label{tab:priors}
\,\\ 
    \centering
    \begin{tabular}{c|c|c}
    \hline
     & parameter & prior/value\\  
    \hline 
    \textbf{background-only} & $\Omo$ &  $\mathcal{U}[0.01, 0.99]$\\
     & $ H_0\rd \; [\kms]$ &  $\mathcal{U}[3650, 18250]$  \\ 
    \hline 
    \textbf{CMB} & $\ocdm\equiv\Omega_{\rm cdm}h^2$ & $\mathcal{U}[0.001, 0.99]$ \\
    & $\ob\equiv\Omega_{\rm b}h^2$ &  $\mathcal{U}[0.005, 0.1]$ \\
    & $\ln(10^{10} A_{s})$ &  $\mathcal{U}[1.61, 3.91]$ \\
    & $n_{s}$ &  $\mathcal{U}[0.8, 1.2]$ \\
    & $H_{0} \; [\kmsMpc]$ &  $\mathcal{U}[20, 100]$  \\
    & $\tau$  &  $\mathcal{U}[0.01, 0.8]$  \\
    \hline 
    \hline \textbf{Hyperparameters} &      $C_{i=0,1,2,3}^{w}$ & $\mathcal{N}(\mu^{\Lambda\rm CDM},\sigma =3)$\\ 
     & $C_{i=0,1,2,3}^{f_{\rm DE}}$ & $\mathcal{N}(\mu^{\Lambda\rm CDM},\sigma =1)$\\
    \hline
    \end{tabular}
\end{table}

One important thing to mention at this point is the choice of prior for the Chebyshev coefficients $C_i$'s (or hyperparameters). 
Similar to what happens in many other analyses, having no ``physically-motivated'' priors for the $C_i$'s can lead to prior-volume effects that could potentially bias the results, as explained in \cref{sec:hyperapx}. To prevent this from happening, we give Gaussian priors on $C_i\sim\mathcal{N}(\mu^{\Lambda\rm CDM},\sigma^2)$, where $\mu^{\Lambda\rm CDM}$ correspond to the point $C_0=1$ and $C_{i>0}=0$. This ensures that any significant deviation from the mean function (\lcdm) is purely driven by the data.

\section{Results}\label{sec:results}

In this section, we present our findings using different data combinations and compare the reconstructions when using crossing statistics on the equation of state $w(z)$, or the energy density $\fde(z)$.

\subsection{Results using \tpdf{$w(z)$}}\label{sec:results_w}
\begin{figure}[th] 
    \centering    
    \includegraphics[width=\textwidth]{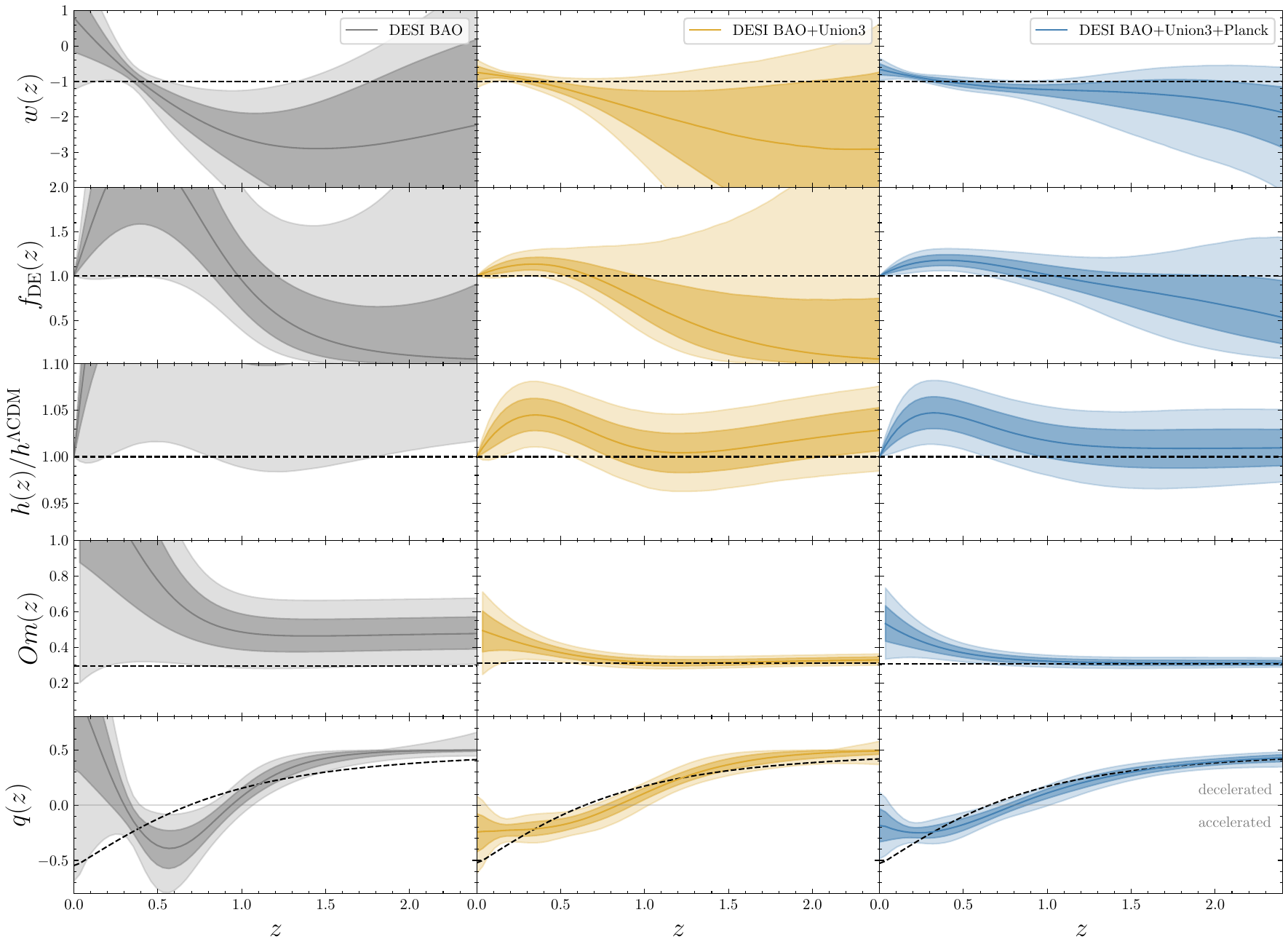}
    \caption{Dark energy reconstructions using the Chebyshev expansion of $w(z)$ up to four terms with DESI BAO, DESI BAO+Union3, and DESI BAO+Union3+Planck datasets, respectively. The colored lines correspond to the median of the posterior distributions and the shaded regions show the $68\%$ and $95\%$ confidence intervals around it. The black-dashed lines depict the best-fit \lcdm\ predictions for each data combination. 
    } 
    \label{fig:DE-w-recos-Union3}
\end{figure}

We start by discussing the results when treating the dark energy equation of state parameter $w(z)$ as a Chebyshev series up to four terms. In the top panels of \cref{fig:DE-w-recos-Union3}, we show the redshift evolution of the equation of state, $w(z)$, while the second row shows its corresponding effective dark energy density, $\fde(z)$, for various data combinations. The third row shows the constraints on the shape of the expansion history, $h(z)=H(z)/H_0$, normalized to the best fit \lcdm\ model for visual clarity. In the fourth row, we show the corresponding $Om(z)$ diagnostic.
The $Om(z)$ diagnostic \cite{Sahni:2008xx} was specifically tailored to efficiently distinguish dark energy models from a cosmological constant, and is defined as 
\begin{equation}
    Om(z)\equiv\frac{h^2(z)-1}{(1+z)^3 - 1}~.
\end{equation}
For $\Lambda$CDM one has simply $Om(z)=\Omo$. Finally, in the bottom panel, we also plot the evolution of the deceleration parameter $q(z)$, defined as
\begin{equation}
    q(z)\equiv-\frac{\ddot{a}a}{\dot{a}^2}= -\frac{\dot H}{H^2}-1=\frac{d \ln{H}}{d\ln{(1+z)}}-1~.
\end{equation}

In all cases, the dashed black lines represent the \lcdm\ best-fit values for these quantities, and for each data combination. 
The coloured lines correspond to the median of the posterior distributions and the shaded regions show the $68\%$ and $95\%$ confidence intervals around it, respectively. 

\begin{figure}[t] 
    \centering
\includegraphics[width=0.47\textwidth]{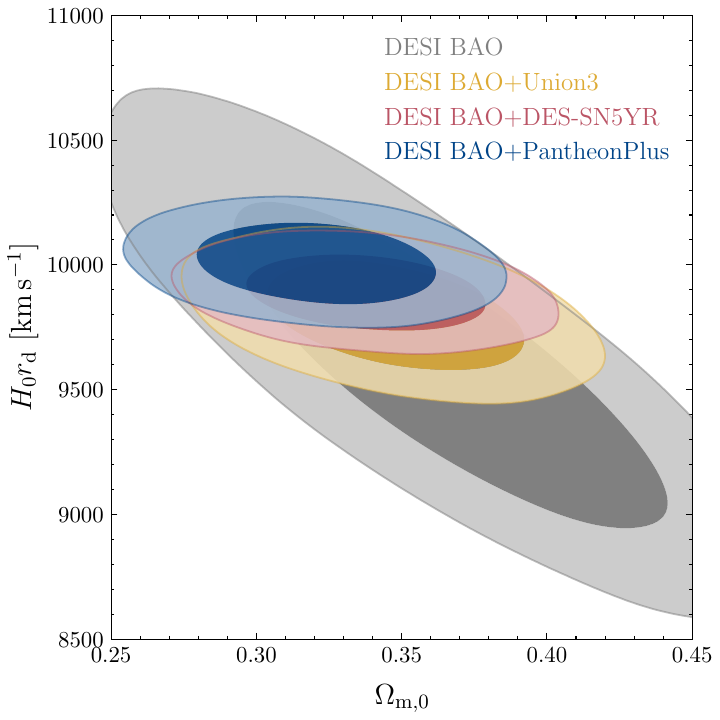} 
\quad 
\includegraphics[width=0.47\textwidth]{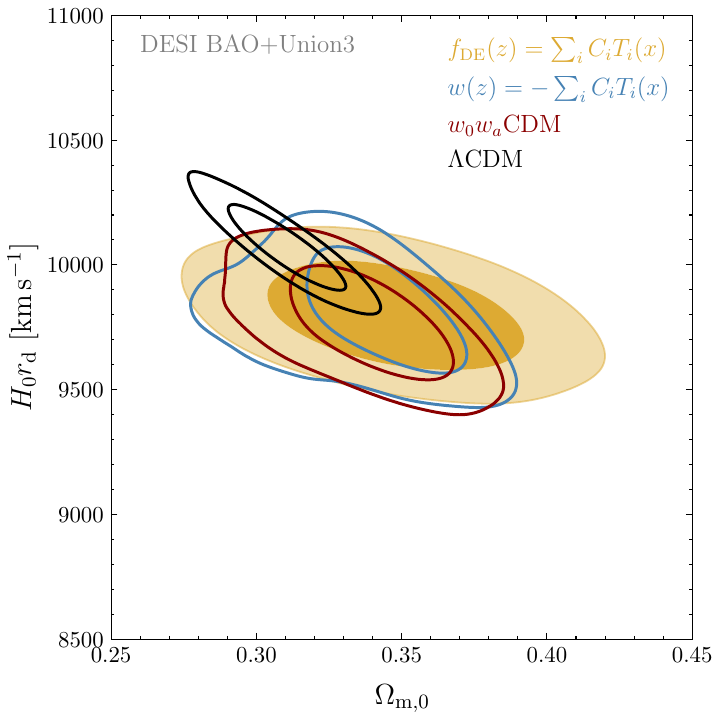} 
    \caption{Marginalized posterior distributions for the cosmological parameters $H_0 r_d$ and $\Omo$ using a Chebyshev expansion of $\fde(z)$ from the combination of DESI BAO and different supernova compilations (left). These constraints are independent of the early universe physics. In the right panel, we compare the constraints obtained from the Chebyshev expansion in $\fde(z)$ and $w(z)$ to those obtained from restricting to \lcdm\ (black) or $w_0 w_a$CDM (red), using the DESI BAO+Union3 data combination. 
    } 
     \label{fig:Triangle-DESI+Union3} 
\end{figure}

When allowing for more freedom in the equation of state, DESI BAO data alone cannot constrain the dark energy well, as there is an intrinsic degeneracy between the matter density, the dark energy evolution \citep[e.g.][]{Wasserman:2002gb,PhysRevD.80.123001,Shafieloo:2011zv,von_Marttens_2020,Calderon:2022cfj}, and the absolute scaling set by the combination \Hord\ (shown in \cref{fig:Triangle-DESI+Union3}) \citep{Percival_2007,Cuceu_2019}. These degeneracies lead to very peculiar shapes of $\fde(z)$ that can be compensated by anomalously large matter densities. 
Such dark energy models can fit the DESI data well, the best-fit having a $\dchisq\simeq-5.5$ with respect to \lcdm, while having 4 additional degrees of freedom. 

However, when including distance measurements from SNe Ia these degeneracies are broken by a more accurate determination of $\Omo$ and \Hord, and the dark energy evolution is much more tightly constrained, as shown in the middle column of \cref{fig:DE-w-recos-Union3}. 
The best-fit model from the DESI+Union3 combination leads to an improvement in the fit of $\dchisq\simeq-9.1$ with respect to \lcdm. It is interesting to note that for both DESI BAO and DESI BAO+Union3, a universe that returns to deceleration at present ($q_0\geq0$) is consistent with the data. At higher redshifts, the difference in $q(z\gtrsim1)$ relative to \lcdm\ reflects the preference for higher matter densities with respect to \lcdm\ (also seen in $Om(z\gtrsim1))$, which implies a longer epoch of matter domination ($q=1/2$). In the rightmost panels, we show the reconstructions when including the measurements of the CMB anisotropies by Planck. These measurements probe physics at much higher redshifts ($\zrec\simeq1100$), and provide a better estimation of the physical matter density $\omega_{\mathrm m}=\Omo h^2$, as the height and position of the acoustic peaks are very sensitive to $\ocdm$ and $\ob$.
The spacing between these peaks is exquisitely measured by Planck \cite{Planck:2018vyg}, providing constraints on the late-time expansion history through the acoustic scale $\theta_s(\zrec)\propto 1/D_A(\zrec)$, assuming standard pre-recombination physics. The CMB constraints, in combination with lower redshift SN and BAO measurements, lead to a much smoother reconstructed dark energy behaviour and a corresponding expansion history that mimics closely that of \lcdm\ at higher redshifts. Our best-fit reconstruction leads to an improvement in the fit corresponding to a $\dchisq\simeq-14.6$ compared to \lcdm.

\begin{table}[ht]
    \centering
     \caption{Constraints on $\Omo$ and \Hord\ using the Chebyshev expansion up to four terms in the equation of state $w(z)$ and different non-CMB dataset combinations. 
     } 
        \label{tab:Om0-Hrd}   
    $\,$\\ 
        \begin{tabular}{ccc} \hline\hline  
         Data&  $\Omo$& \Hord~[$\kms$]\ \\ \hline  
         DESI BAO+Union3& $0.339^{+0.022}_{-0.015}$  & $9808\pm 150$\\ 
         DESI BAO+DES-SN5YR& $0.328^{+0.026}_{-0.011}$ & $9891\pm 100$\\ 
         DESI BAO+PantheonPlus& $0.327^{+0.017}_{-0.012}$ & $10030\pm 110$\\ \hline\hline 
    \end{tabular}
\end{table}

Moreover, combining BAO with SNe~Ia measurements allows for a (relatively) model-independent determination of $\Hord$, with no assumptions on the physics of the early Universe and for an extended class of dark energy models. In \cref{tab:Om0-Hrd}, we present the constraints on $\Hord$ and $\Omo$ from various data combinations, also shown as 2D distributions in the right panel of \cref{fig:Triangle-DESI+Union3}. Note that while our analysis was carried out using a much more flexible parametrization for the dark energy,
the inferred values of \Omo\ and $H_0r_d$ are consistent with those reported in \cite{DESI2024.VI.KP7A} in \lcdm\ and $w_0w_a\rm CDM$, as seen from the right panel of \cref{fig:Triangle-DESI+Union3}

Our methodology provides us with an extended class of expansion histories that are consistent with the data, and that lead to an improvement in fit. In \cref{tab:best-fit}, we report the $\dchisq$ values for the different data combinations, relative to the best-fit $\chisq$ values in \lcdm. 
Note that the results are quite consistent with the $w_0$--$w_a$ approach, with crossing statistics having two more parameters than $w_0$--$w_a$ but comparable $\chisq$. Both approaches however have noticeably better fits than \lcdm. Thus our reconstruction results with more freedom support the conclusions obtained using the standard $w_0$--$w_a$ in \cite{DESI2024.VI.KP7A}. Furthermore, the more probes combined, the greater favouring of these models over \lcdm. 
Irrespective of the dataset combination, the preference for a vanishing, or at least diminished, dark energy component in the past remains. Finally, let us note that due to the wide plotting range chosen for the equation of state ($-4<w<1$), the deviations from a cosmological constant, $w=-1$, may not look visually substantial. However, these deviations from \lcdm\ are better reflected in the expansion history $h(z)$ and 
$Om(z)$ diagnostic (as seen from the third and fourth rows in \cref{fig:DE-w-recos-Union3}).

\begin{table}[ht]
    \centering
    \caption{$\dchisq_{\rm MAP}\equiv \chisq_{\rm model}-\chisq_{\Lambda\rm CDM}$ maximum a posteriori (MAP) values for the different models and data combinations. $\chisq_{w(z)}$ refer to the runs using  $w(z)$, while $\chisq_{f_{\rm DE}}$ refers to the modeling of $\fde(z)$ using a Chebyshev expansion with four terms. The minimum $\chisq$ values were obtained using the minimizers \texttt{iminuit} \cite{iminuit} and \texttt{Py-BOBYQA} \cite{cartis2018improving,Cartis_2021}. Note that all data combinations include DESI BAO measurements. (Recall $\fde(z)$ analysis did not include CMB.)}  
$\,$\\ 

\label{tab:best-fit}
    \begin{tabular}{c|c|c|c|c} \hline  \hline 
         Data & $\Delta \chi^2_{w(z)}$ &$\Delta \chi^2_{\fde}$ &  $\Delta \chi^2_{w_0w_a}$& $\chi^2_{\Lambda\rm CDM}$\\ \hline  
         DESI BAO& $-5.5$ & $-2.1$  & $-3.7$ & $12.7$\\  
         \hline 
         +Union3& $-9.1$& $-8.7$ & $-9.0$& $41.0$\\ 
         +DES-SN5YR& $-11.3$ & $-10.8$ & $-11.3$ & $1659.7$\\ 
         +PantheonPlus& $-5.4$ & $-4.1$ & $-3.6$ & $1418.7$\\ \hline 
         +Union3+Planck& $-14.6$ & - & $-15.7$ & $2810.9$\\ 
         +DES-SN5YR+Planck& $-17.9$ & - & $-19.1$ & $4430.7$  \\ 
         +PantheonPlus+Planck& $-11.1$ & - & $-7.4$ & $4188.4$\\ \hline  \hline 
    \end{tabular}

\end{table}

\subsection{Results using \tpdf{$\fde(z)$}}\label{sec:negative_fde}
\begin{figure}[ht]
    \centering
    \includegraphics[width=\textwidth]{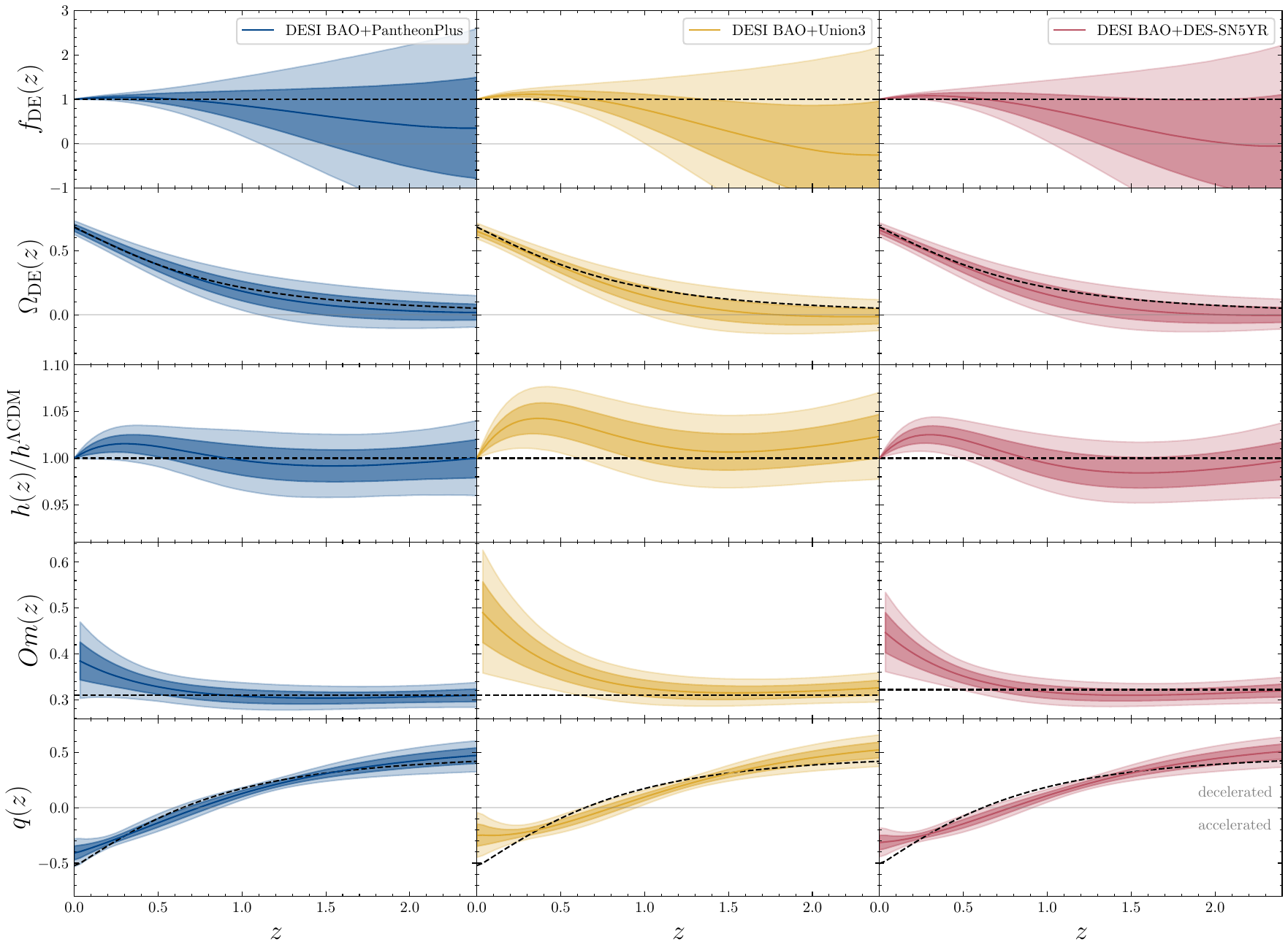}
    \caption{Dark energy reconstructions using the Chebyshev expansion of $\fde(z)$ up to four terms with DESI BAO+Union3, DESI BAO+PantheonPlus and DESI BAO+DES-SN5YR datasets, respectively. The colored lines correspond to the median of the posterior distributions and the shaded regions show the $68\%$ and $95\%$ confidence intervals around it. The black-dashed lines depict the best-fit \lcdm\ predictions for each data combination.}
    \label{fig:negative-fDE-SN+BAO}
\end{figure}

We now turn our attention to the dark energy density. We assess the robustness of our conclusions by allowing for the \textit{effective} energy density to change sign and become negative. These behaviours are not possible to achieve by modeling the dark energy via its equation of state, as it inherently imposes $\fde>0$, as is evident from \cref{eqn_fde}. In contrast with the previous section where we presented results with the Chebyshev expansion of $w(z)$, the direct reconstructions using Chebyshev expansion of $\fde(z)$ are shown in \cref{fig:negative-fDE-SN+BAO}, where we plot the same quantities as in \cref{fig:DE-w-recos-Union3}, replacing $w(z)$ with $\Omega_{\rm DE}(z)$ in the second row since the equation of state becomes singular 
when $\fde$ crosses zero. The overall conclusions drawn from the previous \cref{sec:results_w} are unchanged. One notable difference is that the redshift dependence of $\fde$ is much smoother. However, as discussed in \cref{sec:fde-model}, $C_0$ is no longer a free parameter but is derived by imposing $\fde(z=0)=1$. Thus, strictly speaking, the modeling of $\fde$ has one less degree of freedom compared to that of $w(z)$. In the third column of \cref{tab:best-fit}, we report the $\dchisq$ values for the $\fde(z)$ reconstructions, relative to the best-fit \lcdm. Note that both the $w(z)$ and $\fde(z)$ reconstructions lead to comparable $\chisq$ values for the combined BAO+SNe data, improving over \lcdm.\\

It is interesting to note that the data indeed allows the effective energy density to become negative at $z\gtrsim1$, with the Union3 and DES-SN5YR compilations driving the preference for slightly more negative values with respect to PantheonPlus. Although our approach at this stage is purely phenomenological, negative energy densities can be achieved in various theoretically motivated scenarios, such as modified gravity \citep{Sahni_2003,Zhou_2009,Bauer_2010}, or by invoking the presence of a negative cosmological constant, with an additional degree of freedom driving the accelerated expansion at late-times \cite{Grande:2006nn,Visinelli:2019qqu,Calderon:2020hoc}. As discussed in e.g.\ \citep{Carroll:2004de, PhysRevD.73.083509}, a change of sign in $\rho_{\rm DE}^{\rm eff}(z)$ could indicate non-trivial interactions in the dark sector.\\

While the reconstructions of $\fde(z)$ seem consistent with a cosmological constant at higher redshifts, the reconstructed expansion history $h(z)$ at low redshift shows a $>2\sigma$ deviation from the best-fit \lcdm\ expansion history for each data combination, 
in part due to the different values of $\Omo$. The deviations in $h(z)$ are more prominent since these not only reflect the deviations in $\fde(z)\neq1$, but also the deviations in $\Omo\neq\Omo^{\Lambda\rm CDM}$.
These deviations also (necessarily) translate into $\gtrsim2\sigma$ deviations in the $Om(z\lesssim0.5)$ diagnostic and deceleration parameter, $q(z\lesssim0.3)$. Finally, in \cref{tab:Om0-Hrd-negative-fde} we report the marginalized constraints on the matter density $\Omo$ and $\Hord$ when allowing for negative energy densities.
Note that DESI data alone does not have low redshift measurements and this allows flexible forms of expansion histories, such as those considered in this paper, to vary widely at very low redshifts, and this results in broad and weak constraints on $\Hord$. 
Combined with supernova data, which has some measurements at low redshifts, tightens the constraints on $\Hord$ as seen from \cref{fig:Triangle-DESI+Union3}.

\begin{table}[htbp]
    \centering
    \caption{Constraints on $\Omo$ and \Hord\ using the Chebyshev expansion up to four terms in the dark energy density $\fde(z)$, for different non-CMB dataset combinations. 
    } 
    $\,$\\ 
    \label{tab:Om0-Hrd-negative-fde}
    \begin{tabular}{ccc} \hline\hline  
         Data&  $\Omo$& \Hord~[$\kms$]\ \\ \hline  
         DESI BAO+Union3& $0.347\pm 0.029$ & $9795\pm 140$\\ 
         DESI BAO+DES-SN5YR& $0.338\pm 0.027$ & $9889\pm 100$\\ 
         DESI BAO+PantheonPlus& $0.321\pm 0.027$ & $10010\pm 110$\\ \hline\hline 
    \end{tabular}
\end{table}

\section{Discussion and Conclusions} 

In this paper we use Crossing Statistics, implementing an expansion in terms of Chebyshev polynomials of the equation of state of dark energy  $w(z)$, and dark energy density $\fde(z)$, to reconstruct the expansion history of the universe and properties of dark energy using DESI BAO data combined with supernovae (Union3, PantheonPlus, or DES-SN5YR) and CMB Planck observations. 

Our main result for dark energy is a clear and strong hint towards evolving dark energy, with rapidly falling energy density at $z \gtrsim 1$ (see \cref{fig:DE-w-recos-Union3} middle panels). This behaviour can be modeled by a vanishing dark energy density (going to higher redshifts), with a phantom equation of state $w(z)<-1$ in the recent past. This preference is mainly driven by the non-CMB, ``background-only'' probes. 
Our results also suggest a hump in the expansion rate of the universe at $z\approx0.2$--0.3 relative to the concordance model of cosmology. 
(This is not dependent on our specific choice of $z_{\rm max}$, as shown in Appendix~\ref{sec:zmax} with results little modified when using $z_{\rm max}=2.35$). 

Including CMB observations does not alter this conclusion. However, the transition is smoother, and the expansion history is tightly constrained to mimic that of \lcdm\ at high redshifts, albeit with different dark energy behaviour (see \cref{fig:DE-w-recos-Union3} right panels). This is not unexpected since at high redshifts we do have a matter dominated universe and most models of dark energy (especially phantom ones) would not have much effect on the form of the expansion history. Our findings are consistent with and support the results reported in \cite{DESI2024.VI.KP7A}, showing a very good agreement with the analysis using the $w_0w_a\rm CDM$ parametrization.

\begin{figure}
    \centering
    \includegraphics[width=\textwidth]{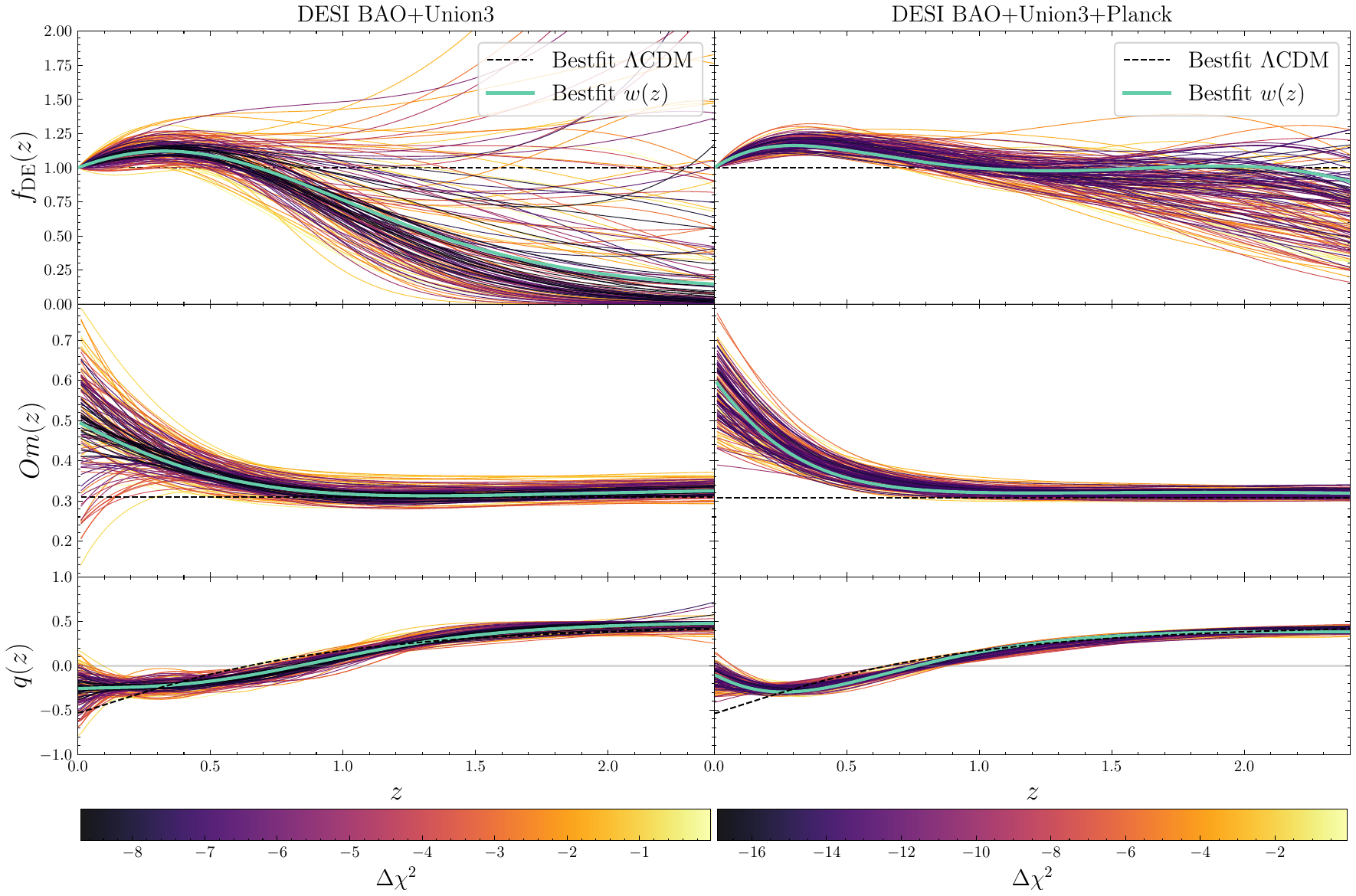}
    \caption{Instead of the marginalized error-bands, here we show some exact reconstructions of the expansion history, in terms of $\fde(z)$, $Om(z)$, and $q(z)$, using the Chebyshev expansion of $w(z)$. Each reconstruction shown has a better likelihood than the best fit standard $\Lambda$CDM model (black dashed lines), color-coded in terms of $\dchisq$, with the green line showing the best fit from the chains.
}\label{fig:omz_q_w_crossing_segments}
\end{figure}

When modeling the dark energy density $\fde(z)$ as a Chebyshev series, and allowing $\fde(z)$ to become negative in some redshift range, we find that the overall trend remains the same, i.e.\ a vanishing dark energy density at $z\gtrsim1.5$ (see \cref{fig:negative-fDE-SN+BAO}). The behaviour of the expansion history from looking at $Om$ diagnostic (being a function of the expansion history alone and sensitive to the combination of matter and dark energy densities) shows clear consistency with the ones obtained using a Chebyshev expansion of $w(z)$.  

The non-CMB probes (SNe~Ia and BAO) allow various shapes for the expansion history but also constrain in a model-independent manner the 
estimation of the absolute scaling set by the combination \Hord, with no assumptions on the physics of the early universe (see \cref{fig:Triangle-DESI+Union3}). 
There is a noticeable shift in the $\Omo-\Hord$ plane between our findings and estimation of these quantities when restricted to the \lcdm\ model fitting Planck and ACT lensing data, quite possibly due in large part to a shift in $\Omo$. 

Our results provide intriguing hints towards evolving dark energy; note that while they were obtained under the assumption of an especially flexible form for $w(z)$, this was still parametric and using any parametric form of $w(z)$ to study the unknown dark energy should be done cautiously. Follow-on work will explore this further. In this paper, we assumed a polynomial expansion (up to four terms, see \cref{exp-order}), which allows the equation of state to take on large negative values at high redshift and hence cause the dark energy density to approach zero. 

While one would ideally like to choose maximally agnostic priors for the crossing hyperparameters, a significantly large part of the hyperparameter space may lead to similar physical quantities, due to the inherent nature of the equation of state (very negative equation of state gives very low energy density and hence little sensitivity). For example, $C_1\gg1$ or $C_3 \gg 1$ results in rapidly vanishing dark energy and similar predictions for the observables. As a result, such cases have very similar physical behaviour and would make posteriors unbounded. To prevent this from biasing our reconstructions, we imposed Gaussian priors on the Chebyshev coefficients centered around \lcdm\ ($C_0=1,C_{i>0}=0$). Such Gaussian priors effectively favour \lcdm\ and ensure that any detected deviation from $w(z)=-1$ is mostly driven by the data.
The marginalized posterior distributions for the Chebyshev coefficients are shown in \cref{sec:hyperapx}. 

To avoid loss of detail by only looking at the marginalized confidence levels, we can also look at the individual reconstructed expansion histories as presented in \cref{fig:omz_q_w_crossing_segments}. These lines are the exact expansion histories having a viable and acceptable fit to the combined data (in fact, with better $\chisq$ than \lcdm). Having similar shapes and trends observed in \cref{fig:omz_q_w_crossing_segments} and \cref{fig:DE-w-recos-Union3} reflects that our choices of parametric form and priors have been reasonable for our case of study. \\

We emphasize that many of the aspects of intriguing dark energy behaviour seen here are also present in the standard $w_0$--$w_a$ analysis of \cite{DESI2024.VI.KP7A}, as shown in \cref{sec:w0wa-comparison}. For example, the rise toward $q(z=0)\approx0$ corresponds to the leveling in $H(z)/(1+z)$ at low redshift and the less negative $w_0$ seen there. One can view this work as supporting the robustness of such indications in \cite{DESI2024.VI.KP7A} even when more freedom is allowed in dark energy characteristics. 

Our results also indicate that a broad range of expansion histories, substantially different from a cosmological constant, are viable and consistent with current data combinations. The trend of the reconstructed expansion history, $Om(z)$ diagnostic, and deceleration parameter $q(z)$ (see \cref{fig:omz_q_w_crossing_segments} lower two panels), opens the door to models with a slowing down of cosmic acceleration \cite{Shafieloo:2009ti}. In fact, a universe with $q(z=0) > 0$ -- after an accelerating period where $q(z)<0$ -- can be still consistent with the combined data. Such a temporary nature to cosmic acceleration could have some interesting theoretical implications worth exploring. At higher redshift, $z\gtrsim1$, on the other hand, the trend of the reconstructed form of the dark energy density allows models with emergent dark energy behaviour~\cite{Poulin:2018zxs, Wang:2018fng, Li:2019yem, Li:2020ybr}. 

Further distance data, e.g.\ from DESI three-year measurements or the Nearby Supernova Factory \cite{NearbySupernovaFactory:2022tad}, will allow deeper exploration into such model-independent expansion history reconstructions. Finally, let us note that the purpose of this work is not to perform model selection or to rule out $\Lambda$ at high statistical significance, but to explore the allowed phenomenology of dark energy thoroughly, and compare to the standard $w_0$--$w_a$ parametrization. In a companion paper \cite{DESI:2024kob}, we show that such phenomenology can be reproduced with merely one additional degree of freedom, with some physical motivation.

\acknowledgments

The authors would like to thank Benjamin L'Huillier for valuable discussions. 
We acknowledge the use of the Boltzmann solver \classy~\cite{class_2011arXiv1104.2932L,class_2011JCAP...07..034B} for the computation of theoretical observables, 
\Cobaya\ \cite{Torrado:2020dgo} for the sampling and \gd\ \cite{Lewis:2019xzd} for the post-processing of our results. We also acknowledge the use of the standard \texttt{python} libraries for scientific computing, such as \texttt{numpy} \cite{harris2020array}, \texttt{scipy} \cite{2020SciPy-NMeth} and \texttt{matplotlib} \cite{Hunter:2007}.
This work was supported by the
high-performance computing cluster Seondeok at the Korea Astronomy and Space Science Institute. A.S. would like to acknowledge the support by National Research Foundation of Korea 2021M3F7A1082056, 
and the support of the Korea Institute for Advanced
Study (KIAS) grant funded by the government of Korea.\\

This material is based upon work supported by the U.S. Department of Energy (DOE), Office of Science, Office of High-Energy Physics, under Contract No. DE–AC02–05CH11231, and by the National Energy Research Scientific Computing Center, a DOE Office of Science User Facility under the same contract. Additional support for DESI was provided by the U.S. National Science Foundation (NSF), Division of Astronomical Sciences under Contract No. AST-0950945 to the NSF’s National Optical-Infrared Astronomy Research Laboratory; the Science and Technology Facilities Council of the United Kingdom; the Gordon and Betty Moore Foundation; the Heising-Simons Foundation; the French Alternative Energies and Atomic Energy Commission (CEA); the National Council of Humanities, Science and Technology of Mexico (CONAHCYT); the Ministry of Science and Innovation of Spain (MICINN), and by the DESI Member Institutions: \url{https://www.desi.lbl.gov/collaborating-institutions}. Any opinions, findings, and conclusions or recommendations expressed in this material are those of the author(s) and do not necessarily reflect the views of the U. S. National Science Foundation, the U. S. Department of Energy, or any of the listed funding agencies. 

The DESI collaboration is honored to be permitted to conduct scientific research on Iolkam Du’ag (Kitt Peak), a mountain with particular significance to the Tohono O’odham Nation.

\section*{Data Availability}


The data used in this analysis will be made public along the Data Release 1 (details in \url{https://data.desi.lbl.gov/doc/releases/}).

\appendix

\section{Order of the Chebyshev expansion}\label{exp-order} 

\begin{figure}
    \centering
\includegraphics[width=\textwidth]{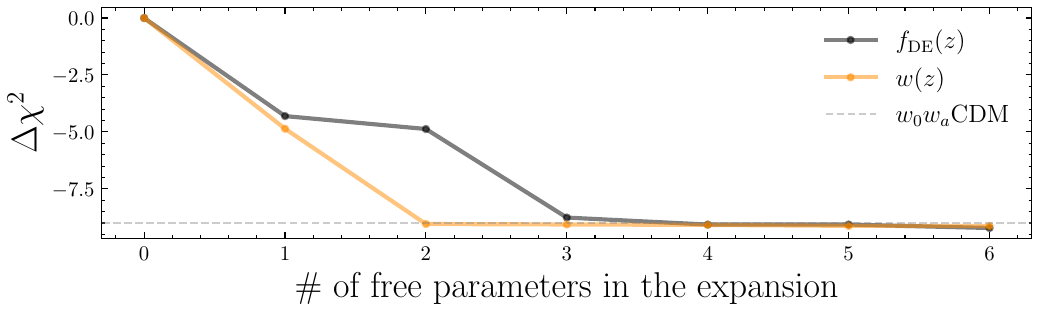}
    \caption{Change in the $\dchisq\equiv\chisq_{\rm cross}-\chisq_{\Lambda\rm CDM}$ value as a function of the number of free parameters introduced in the Chebyshev expansion, for the data combination DESI BAO+Union3.
    }
    \label{fig:chi2_free_params}
\end{figure}

To justify our choice of truncating the Chebyshev series at four terms, in \cref{fig:chi2_free_params} we inspect the behaviour of the \dchisq\ as a function of the number of free parameters for the DESI BAO+Union3 combination. While the expansion in $w(z)$ requires only 2 free parameters for the best-fit \chisq\ to ``converge'' (as known from the success of $w_0$--$w_a$ \citep[see also][]{Linder:2005ne}), the expansion of $\fde(z)$ requires at least 3 free parameters to achieve similar performance. This also motivates why the $\dchisq$'s reported in \cref{tab:best-fit} are similar for $w(z)$,
$\fde(z)$, and $w_0w_a$CDM. 

Note that 3 free parameters in the expansion of $\fde(z)$ correspond to the four $C_i$ in \cref{eq:fCi} since $C_0$ is determined by \cref{eq:C0}. This is sufficient to ensure a similar performance between the $w(z)$ and $\fde(z)$ modeling approaches, and a fair comparison of the different reconstructions. Thus, we restrict ourselves to a four-term Chebyshev expansion, after which the \chisq\ does not change significantly by adding more degrees of freedom. This also matches previous findings in the context of CMB analyses \cite{Hazra:2014hma}.

\section{Robustness of \tpdf{$z_{\rm max}$}}\label{sec:zmax}

\begin{figure}[h]
    \centering
    \includegraphics[width=\textwidth]{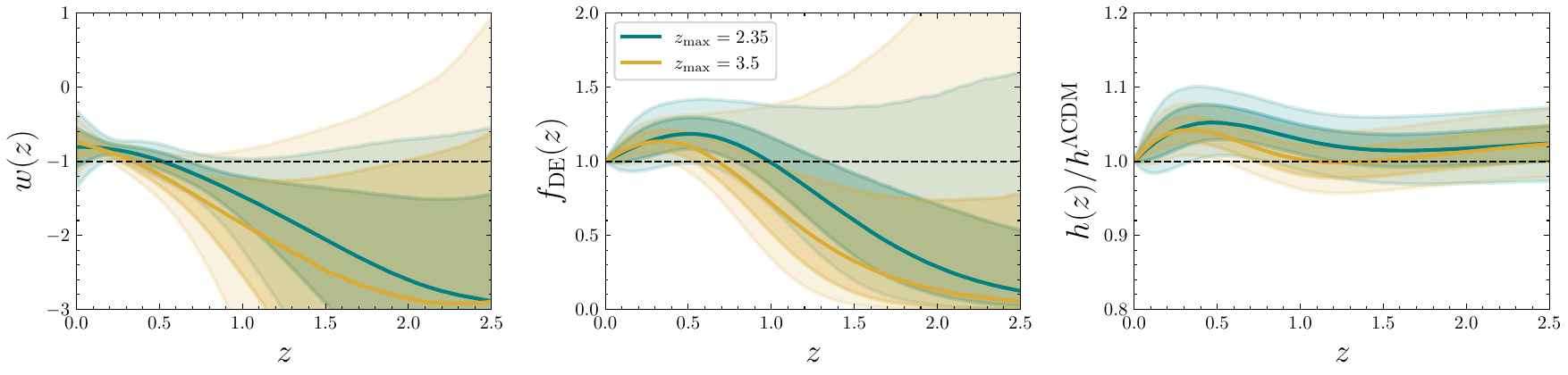}
    \caption{Effect of $z_{\rm max}$ on the dark energy reconstructions with the $w(z)$ crossing approach, shown for DESI BAO+Union3. 
}
    \label{fig:w_recon_zmax_effect}
\end{figure}

\begin{figure}[h]
    \centering
    \includegraphics[width=0.45\textwidth]{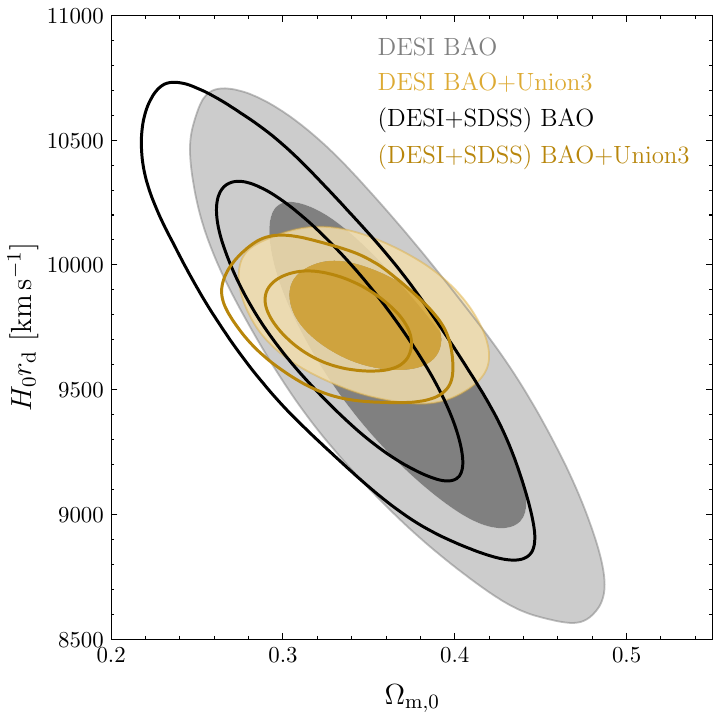}
    \includegraphics[width=0.45\textwidth]{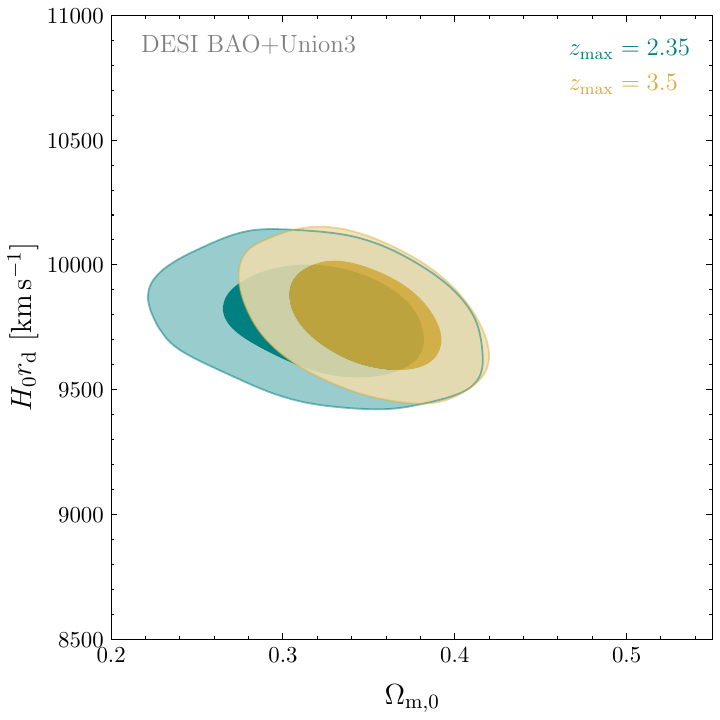}
    \caption{We compare the marginalized constraints on $\Hord$  and $\Omo$ when using DESI BAO vs (DESI+SDSS) BAO data (left panel) or the two different values of $z_{\rm max}$ (right panel), for the $\fde(z)$ crossing approach. 
    }
    \label{fig:cosmo_recon_zmax_effect}
\end{figure}

To check that the results do not depend to any significant degree on the value of $z_{\rm max}$, we show results in \cref{fig:w_recon_zmax_effect} that compare 
those using $z_{\rm max}=2.35$ to our standard $z_{\rm max}=3.5$. The general features of a strongly negative $w(z)$ at $z\gtrsim0.7$, and hence vanishing dark energy density, and a rise in $w(z)$ at $z\lesssim0.2$, are robust; these propagate to close similarity for $\fde(z)$ and $h(z)$ as well. 
The conclusions on the matter density $\Omo$ and $\Hord$ also remain substantially unaffected, as seen in \cref{fig:cosmo_recon_zmax_effect}.

\section{Crossing hyperparameters and effect of priors} \label{sec:hyperapx} 

\begin{figure}[h!]
    \centering
    \includegraphics[width=\textwidth]{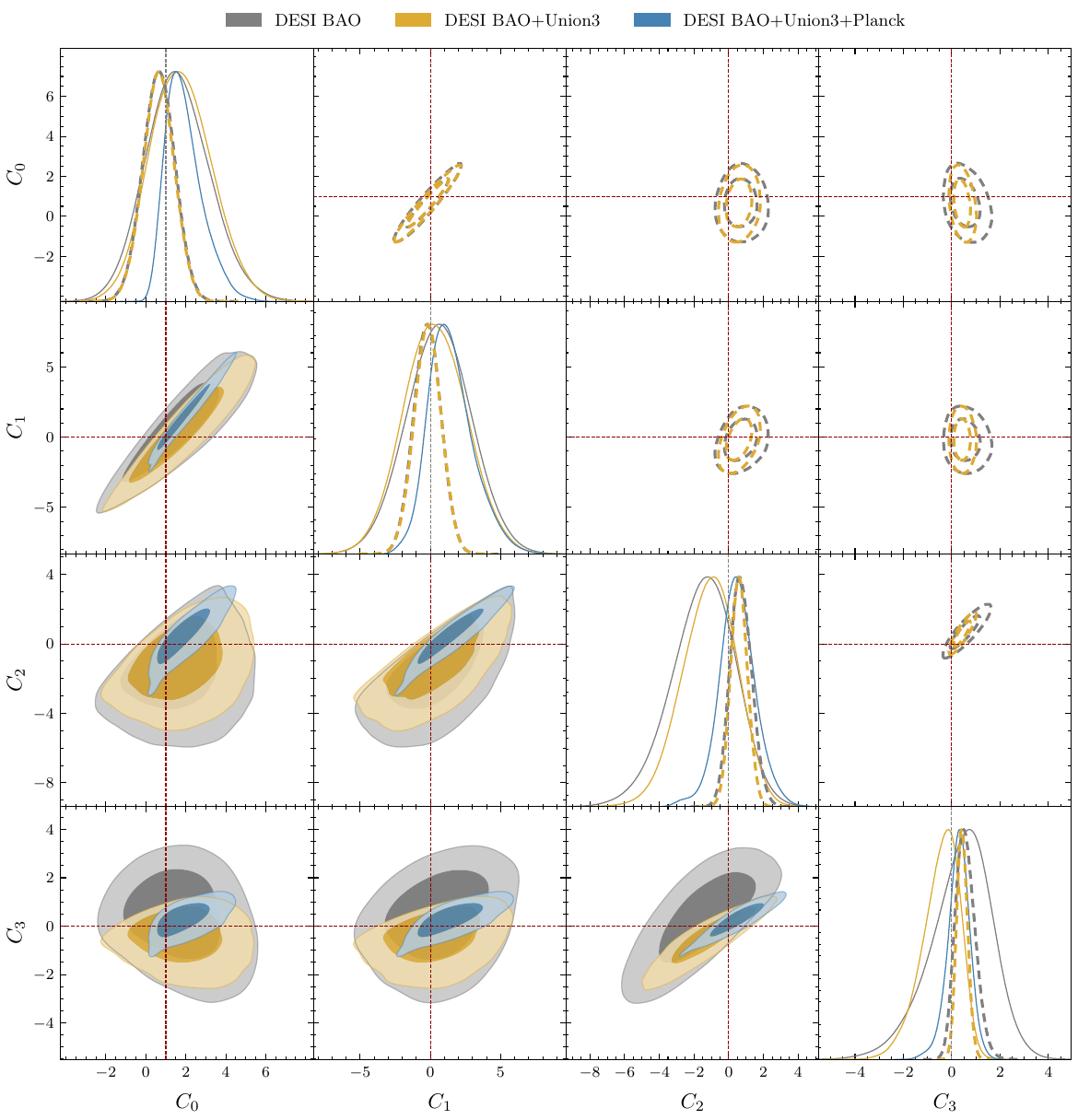}
    \caption{Marginalized posterior distributions on the Chebyshev coefficients for the $w(z)$ (solid contours, lower triangle) and the $\fde(z)$ (dashed contours, upper triangle) Chebyshev expansions. The red-dashed lines give the values taken by the mean function (\lcdm). 
    }

    \label{fig:crossing_hyperparameters}
\end{figure}

We present the posterior distribution for Chebyshev polynomial coefficients for $w(z)$ (lower triangle) and $\fde(z)$ (upper triangle) in \cref{fig:crossing_hyperparameters} utilising the Gaussian priors from \cref{tab:priors}. Note that certain areas of the hyperparameter space produce comparable observables. This is especially evident in \cref{eqn_fde,eq:h2} when $C_i$ takes high values that cause $\fde(z)$ to drop rapidly at higher redshift. Such vanishing dark energy yields indiscernible predictions for the observables ($\DMrd,\DHrd$, and $\DVrd$). This results in a large number of models with similar likelihood values and unbounded posterior distributions for the coefficients, unless the data has enough constraining power to break the $\Omo-\fde$ degeneracy. (The same effect is evident in \cite{DESI2024.VI.KP7A} for $w_0w_a$CDM.) 
Thus, having a large number of hyperparameter combinations that result in similar observables can introduce unwanted biases in the marginalized posterior distributions of $w(z)/\fde(z)$. To minimize potential prior-volume effects, we adopt Gaussian priors on the Chebyshev coefficients centered around the \lcdm\ expected values $C_0=1$ and $C_{i>0}=0$. This is, in fact, a conservative choice, as we are effectively penalizing large deviations from $C_0=1$ and $C_{i>0}=0$, and intentionally favouring \lcdm. Nonetheless, our choice of the Gaussian prior width is large enough that it allows us to cover a wide range of dark energy behaviours, as seen by the posterior distributions in \cref{fig:DE-w-recos-Union3,fig:negative-fDE-SN+BAO}. Our findings (see the $\Delta\chi^2$ in \cref{tab:best-fit}) indicate that despite having Gaussian priors on the $C_i$'s, the data still favours deviations from \lcdm, adding to the robustness of our results. 

\section{Comparison with (DESI+SDSS)}\label{sec:SDSS} 

\begin{figure}[h!]
    \centering
    \includegraphics[width=\textwidth]{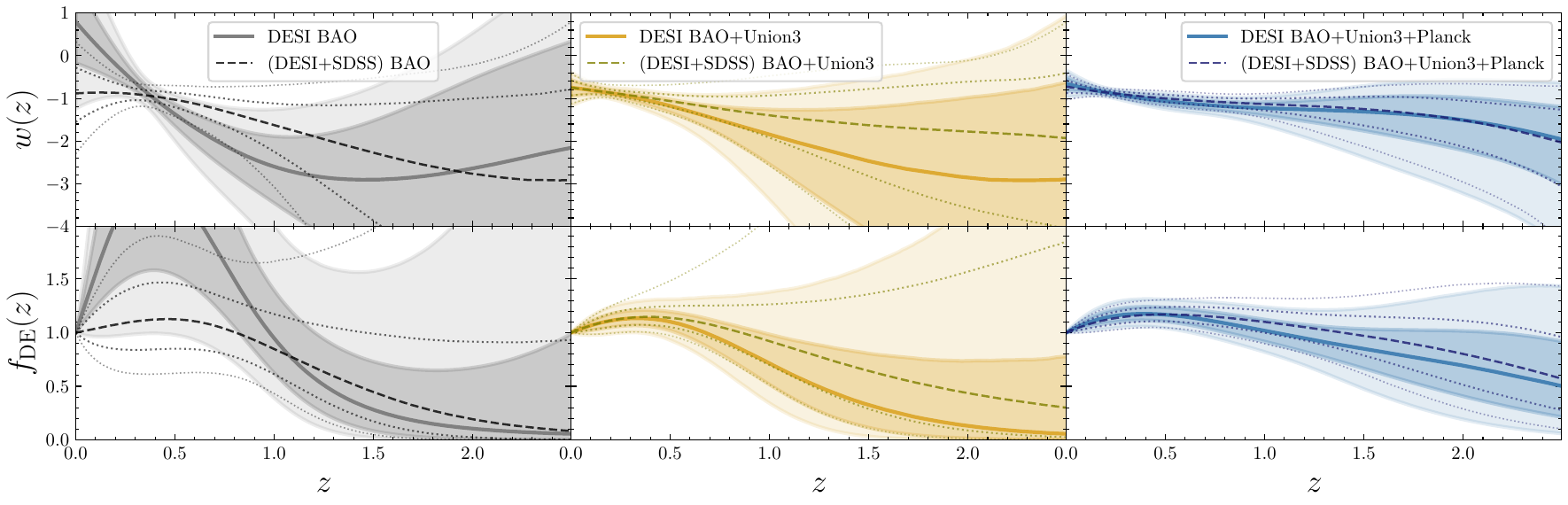} 
    \caption{Comparison of the dark energy reconstructions using the Chebyshev expansion of $w(z)$ from (DESI+SDSS) BAO (dashed lines) vs DESI BAO (solid-filled contours).}
    \label{fig:recos-fde-sdss-comp}
\end{figure}

\Cref{fig:recos-fde-sdss-comp} compares the results when using DESI BAO (as in the main text) vs the combination of (DESI+SDSS) BAO (as described in \cref{sec:data} and in detail in \cite{DESI2024.VI.KP7A}, this is not a mere addition of the two sets but a combination of the most impactful points from both sets). The results are quite consistent, in particular when considering the full DESI BAO+Union3+Planck data combination. 
The conclusions on the matter density $\Omo$ and $\Hord$ also remain substantially unaffected, as was seen in \cref{fig:cosmo_recon_zmax_effect}. We refer the reader to Appendix A in \cite{DESI2024.VI.KP7A} for a more thorough comparison of the (DESI+SDSS) vs DESI BAO results.

\section{Comparison with \tpdf{$w_0w_a\rm CDM$}}\label{sec:w0wa-comparison}
\begin{figure}[h!]
    \centering
    \includegraphics[width=\textwidth]{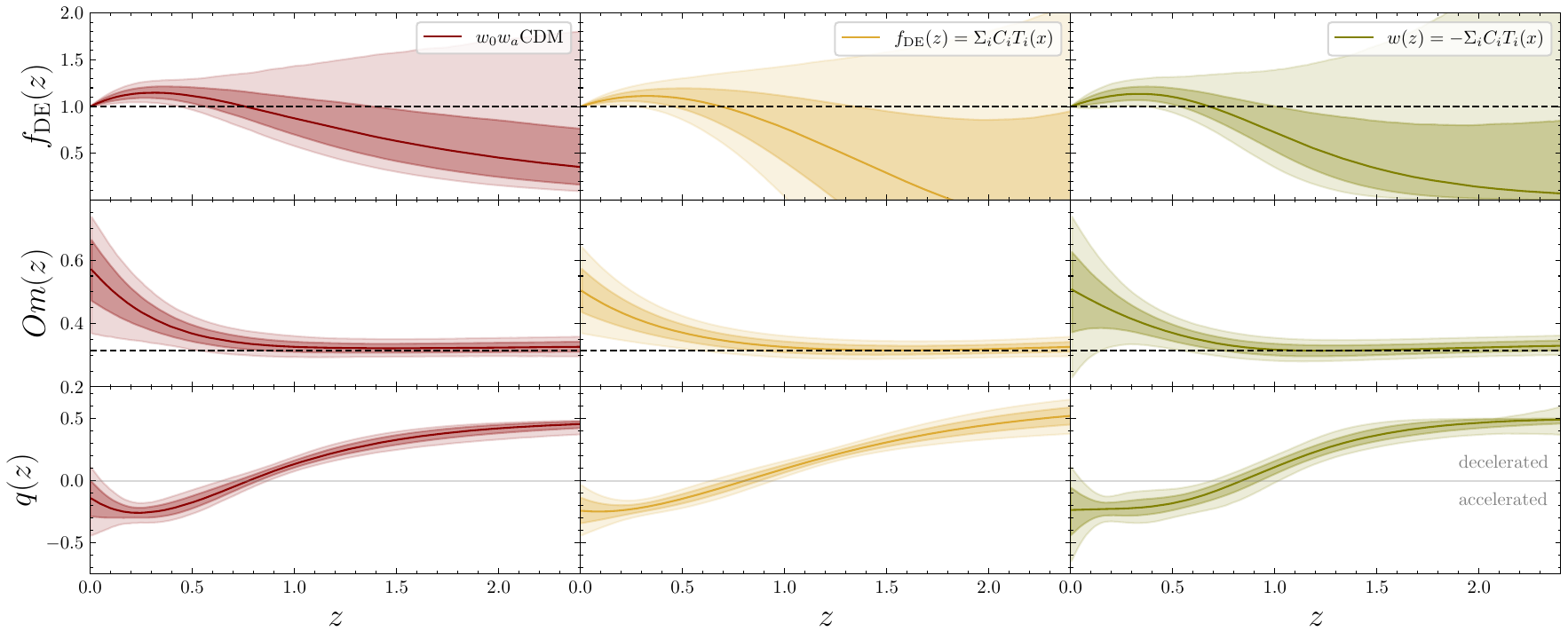}
\caption{Comparison of the dark energy reconstructions using the conventional $w_0$--$w_a$ parametrization, a Chebyshev expansion of $\fde(z)$ and a Chebyshev expansion of $w(z)$ from using DESI BAO+Union3 data.}
    \label{fig:comparison-w0wa}
\end{figure}

In \Cref{fig:comparison-w0wa}, we show how our results compare to the results using the conventional $w_0w_a\rm CDM$ parametrization.

\bibliographystyle{JHEP}
\bibliography{biblio.bib,DESI2024_Supporting_Articles}


\section{Author Affiliations}
\label{sec:affiliations}

\noindent \hangindent=.5cm $^{1}${Korea Astronomy and Space Science Institute, 776, Daedeokdae-ro, Yuseong-gu, Daejeon 34055, Republic of Korea}

\noindent \hangindent=.5cm $^{2}${University of Science and Technology, 217 Gajeong-ro, Yuseong-gu, Daejeon 34113, Republic of Korea}

\noindent \hangindent=.5cm $^{3}${Lawrence Berkeley National Laboratory, 1 Cyclotron Road, Berkeley, CA 94720, USA}

\noindent \hangindent=.5cm $^{4}${Space Sciences Laboratory, University of California, Berkeley, 7 Gauss Way, Berkeley, CA  94720, USA}

\noindent \hangindent=.5cm $^{5}${University of California, Berkeley, 110 Sproul Hall \#5800 Berkeley, CA 94720, USA}

\noindent \hangindent=.5cm $^{6}${IRFU, CEA, Universit\'{e} Paris-Saclay, F-91191 Gif-sur-Yvette, France}

\noindent \hangindent=.5cm $^{7}${Departamento de F\'{i}sica, Instituto Nacional de Investigaciones Nucleares, Carreterra M\'{e}xico-Toluca S/N, La Marquesa,  Ocoyoacac, Edo. de M\'{e}xico C.P. 52750,  M\'{e}xico}

\noindent \hangindent=.5cm $^{8}${Institute of Cosmology and Gravitation, University of Portsmouth, Dennis Sciama Building, Portsmouth, PO1 3FX, UK}

\noindent \hangindent=.5cm $^{9}${School of Mathematics and Physics, University of Queensland, 4072, Australia}

\noindent \hangindent=.5cm $^{10}${Department of Physics, The University of Texas at Dallas, Richardson, TX 75080, USA}

\noindent \hangindent=.5cm $^{11}${Departamento de F\'{i}sica, Universidad de Guanajuato - DCI, C.P. 37150, Leon, Guanajuato, M\'{e}xico}

\noindent \hangindent=.5cm $^{12}${Instituto Avanzado de Cosmolog\'{\i}a A.~C., San Marcos 11 - Atenas 202. Magdalena Contreras, 10720. Ciudad de M\'{e}xico, M\'{e}xico}

\noindent \hangindent=.5cm $^{13}${Physics Dept., Boston University, 590 Commonwealth Avenue, Boston, MA 02215, USA}

\noindent \hangindent=.5cm $^{14}${Physics Department, Stanford University, Stanford, CA 93405, USA}

\noindent \hangindent=.5cm $^{15}${SLAC National Accelerator Laboratory, Menlo Park, CA 94305, USA}

\noindent \hangindent=.5cm $^{16}${Department of Physics \& Astronomy, University College London, Gower Street, London, WC1E 6BT, UK}

\noindent \hangindent=.5cm $^{17}${Instituto de F\'{\i}sica, Universidad Nacional Aut\'{o}noma de M\'{e}xico,  Cd. de M\'{e}xico  C.P. 04510,  M\'{e}xico}

\noindent \hangindent=.5cm $^{18}${NSF NOIRLab, 950 N. Cherry Ave., Tucson, AZ 85719, USA}

\noindent \hangindent=.5cm $^{19}${Department of Physics \& Astronomy and Pittsburgh Particle Physics, Astrophysics, and Cosmology Center (PITT PACC), University of Pittsburgh, 3941 O'Hara Street, Pittsburgh, PA 15260, USA}

\noindent \hangindent=.5cm $^{20}${Departamento de F\'isica, Universidad de los Andes, Cra. 1 No. 18A-10, Edificio Ip, CP 111711, Bogot\'a, Colombia}

\noindent \hangindent=.5cm $^{21}${Observatorio Astron\'omico, Universidad de los Andes, Cra. 1 No. 18A-10, Edificio H, CP 111711 Bogot\'a, Colombia}

\noindent \hangindent=.5cm $^{22}${Institut d'Estudis Espacials de Catalunya (IEEC), 08034 Barcelona, Spain}

\noindent \hangindent=.5cm $^{23}${Institute of Space Sciences, ICE-CSIC, Campus UAB, Carrer de Can Magrans s/n, 08913 Bellaterra, Barcelona, Spain}

\noindent \hangindent=.5cm $^{24}${Center for Cosmology and AstroParticle Physics, The Ohio State University, 191 West Woodruff Avenue, Columbus, OH 43210, USA}

\noindent \hangindent=.5cm $^{25}${Department of Physics, The Ohio State University, 191 West Woodruff Avenue, Columbus, OH 43210, USA}

\noindent \hangindent=.5cm $^{26}${The Ohio State University, Columbus, 43210 OH, USA}

\noindent \hangindent=.5cm $^{27}${Sorbonne Universit\'{e}, CNRS/IN2P3, Laboratoire de Physique Nucl\'{e}aire et de Hautes Energies (LPNHE), FR-75005 Paris, France}

\noindent \hangindent=.5cm $^{28}${Departament de F\'{i}sica, Serra H\'{u}nter, Universitat Aut\`{o}noma de Barcelona, 08193 Bellaterra (Barcelona), Spain}

\noindent \hangindent=.5cm $^{29}${Institut de F\'{i}sica d’Altes Energies (IFAE), The Barcelona Institute of Science and Technology, Campus UAB, 08193 Bellaterra Barcelona, Spain}

\noindent \hangindent=.5cm $^{30}${Instituci\'{o} Catalana de Recerca i Estudis Avan\c{c}ats, Passeig de Llu\'{\i}s Companys, 23, 08010 Barcelona, Spain}

\noindent \hangindent=.5cm $^{31}${Department of Physics and Astronomy, Siena College, 515 Loudon Road, Loudonville, NY 12211, USA}

\noindent \hangindent=.5cm $^{32}${Department of Physics and Astronomy, University of Waterloo, 200 University Ave W, Waterloo, ON N2L 3G1, Canada}

\noindent \hangindent=.5cm $^{33}${Perimeter Institute for Theoretical Physics, 31 Caroline St. North, Waterloo, ON N2L 2Y5, Canada}

\noindent \hangindent=.5cm $^{34}${Waterloo Centre for Astrophysics, University of Waterloo, 200 University Ave W, Waterloo, ON N2L 3G1, Canada}

\noindent \hangindent=.5cm $^{35}${Instituto de Astrof\'{i}sica de Andaluc\'{i}a (CSIC), Glorieta de la Astronom\'{i}a, s/n, E-18008 Granada, Spain}

\noindent \hangindent=.5cm $^{36}${Department of Physics, Kansas State University, 116 Cardwell Hall, Manhattan, KS 66506, USA}

\noindent \hangindent=.5cm $^{37}${Department of Physics and Astronomy, Sejong University, Seoul, 143-747, Korea}

\noindent \hangindent=.5cm $^{38}${CIEMAT, Avenida Complutense 40, E-28040 Madrid, Spain}

\noindent \hangindent=.5cm $^{39}${Department of Physics, University of Michigan, Ann Arbor, MI 48109, USA}

\noindent \hangindent=.5cm $^{40}${University of Michigan, Ann Arbor, MI 48109, USA}

\noindent \hangindent=.5cm $^{41}${Department of Physics \& Astronomy, Ohio University, Athens, OH 45701, USA}

\noindent \hangindent=.5cm $^{42}${National Astronomical Observatories, Chinese Academy of Sciences, A20 Datun Rd., Chaoyang District, Beijing, 100012, P.R. China}

\end{document}